\documentclass[onecollarge,natbib]{svjour2}
\bibpunct{[}{]}{;}{n}{}{,} 
\smartqed  

 \usepackage{graphicx,amsmath,amssymb}
\usepackage{indentfirst}
 \usepackage[usenames]{color}
 \usepackage[colorlinks=true, urlcolor=navyblue, linkcolor=navyblue, citecolor=navyblue]{hyperref}
\usepackage{epstopdf}
\usepackage{appendix}

\definecolor{navyblue}{rgb}{0,0.08,0.45}

\def\Dslash{\raise.15ex\hbox{/}\kern-.7em D}
\def\Pslash{\raise.15ex\hbox{/}\kern-.7em P}

\newcommand{\beq}{\begin{equation}}
\newcommand{\enq}{\end{equation}}
\newcommand{\beqa}{\begin{eqnarray}}
\newcommand{\beqast}{\begin{eqnarray*}}
\newcommand{\enqa}{\end{eqnarray}}
\newcommand{\enqast}{\end{eqnarray*}}
\newcommand{\beml}{\begin{multline}}
\newcommand{\enml}{\end{multline}}
\newcommand{\nn}{\nonumber}
\newcommand{\req}[1]{(\ref{#1})}

\newcommand{\bec}{\begin{center}}
\newcommand{\enc}{\end{center}}
\newcommand{\beqo}{\begin{quote}}
\newcommand{\enqo}{\end{quote}}

\newcommand{\half}{{\textstyle{\frac{1}{2}}}}

\newcommand{\mbf}[1]{\mathbf{#1}}

\newcommand{\la}{\lambda}

\journalname{Few Body Systems}
\begin{document}

\title{The Light-Front Schr\"odinger Equation and Determination of the Perturbative QCD Scale from Color Confinement
}
\subtitle{A First Approximation to QCD}


\author{Stanley J. Brodsky$^{a},$\\
Guy F. de T\'eramond$^{b},$\\Alexandre Deur{$^c$},\\ and Hans G\"unter Dosch{$^d$}.}

\institute{\small $^a${\it SLAC National Accelerator Laboratory, Stanford University, Stanford, CA 94309, USA} \\
                            $^b${\it Universidad de Costa Rica, San Jos\'e, Costa Rica} \\
                             $^c${\it Thomas Jefferson National Accelerator Facility, Newport News, VA 23606, USA}\\
                             $^d${\it Institut f\"ur Theoretische Physik, Philosophenweg 16, D-6900 Heidelberg, Germany} }


\date{ }

\maketitle

\begin{abstract}
The  valence Fock-state wavefunctions  of the light-front QCD Hamiltonian satisfy a relativistic equation of motion, analogous to the nonrelativistic radial Schr\"odinger equation, with an effective confining potential $U$ which systematically incorporates the effects of higher quark and gluon Fock states.   
If one requires that the  
effective action  which underlies the  QCD Lagrangian remains conformally invariant and extends the formalism of de Alfaro, Fubini and Furlan to light front Hamiltonian theory, the potential U has a unique form of a harmonic oscillator potential,  and a  mass gap arises.  
The result is a nonperturbative relativistic light-front quantum mechanical wave equation which incorporates color confinement and other essential spectroscopic and dynamical features of hadron physics, including a massless pion for zero quark mass and linear Regge trajectories 
with the same slope in the 
radial quantum number $n$ and orbital angular momentum $L$.   Only one mass parameter $\kappa$ appears. The corresponding light-front Dirac equation provides a dynamical and spectroscopic model of nucleons. The same light-front  equations arise from the holographic mapping 
of  the soft-wall model modification of AdS$_5$ space with a unique dilaton profile  to QCD (3+1) at fixed light-front time.  Light-front holography thus provides a precise relation between the bound-state amplitudes in the fifth dimension of AdS space and the boost-invariant light-front wavefunctions describing the internal structure of hadrons in physical space-time.   
We  also show how the mass scale $\kappa$ underlying confinement and hadron masses  determines the scale   $\Lambda_{\overline{MS}} $ controlling the evolution of the perturbative QCD coupling.  The relation between scales is obtained by matching the nonperturbative dynamics, as described by an effective conformal theory mapped to the light-front and its embedding in AdS space, to the perturbative QCD  regime  computed to four-loop order. The result is an effective coupling  defined at all momenta. The predicted value $\Lambda_{\overline{MS}} = 0.328 \pm 0.034$ GeV is in agreement with the world average $0.339 \pm 0.010$ GeV. The analysis applies to any renormalization scheme.

\end{abstract}

\section{Light-Front Quantization}
\label{intro}

The quantization of QCD at fixed light-front (LF)  time 
$x^+ = x^0 + x^3$ provides a first-principles method for solving nonperturbative QCD.  Given the Lagrangian, one can compute the LF Hamiltonian $H_{LF}$ in terms of the independent quark and gluon fields. The eigenvalues of  $H_{LF}$  determine the mass-squared values of both the discrete and continuum hadronic spectra. The eigensolutions $|\Psi_H\rangle $ projected on the free n-parton Fock state  $\langle n |\Psi_H \rangle$  determine the LF wavefunctions 
$\psi_{n/H}(x_i, \vec k_{\perp i} , \lambda_i) $ where the $x_i = {k^+_i\over P^+} = {k^0_i + k^3_i\over P^0 + P^3}$, with $\sum^n_{i =1} x_i = 1$, are the LF momentum fractions.   The eigenstates  are defined at fixed 
$x^+$  within the causal horizon, so that causality is maintained without normal-ordering.  As Dirac showed ~\cite{Dirac:1949cp}, the front form has the maximum number of kinematic generators of the Lorentz group, including the boost operator, so that the wavefunction is boost invariant. Thus the description of a hadron at fixed 
 $x^+$  is independent of the observer's Lorentz frame, making it the natural formalism for addressing dynamical processes in QCD.  In fact, measurements of hadron structure, such as deep inelastic 
lepton-proton scattering $\ell p \to \ell^\prime X$, determine  the wavefunction and structure of the proton at fixed LF time.    
Light-front physics is a fully relativistic field theory, but its structure is similar to 
non-relativistic atomic physics, and the resulting bound-state equations can be formulated as relativistic Schr\"odinger-like equations at equal LF time. An extensive review is given in Ref.~\cite{Brodsky:1997de}.  Light-front quantization is thus the natural framework for interpreting measurements like deep inelastic scattering in terms of  the nonperturbative relativistic bound-state structure of hadrons in  QCD.  The light-front wavefunctions (LFWFs) of hadrons  provide a direct connection between observables and the QCD Lagrangian.  Moreover, the formalism is rigorous, relativistic, and frame-independent.

Given the frame-independent 
LFWFs $\psi_{n/H}$, one can compute a large range of hadronic
observables, starting with form factors, structure functions, generalized three-dimensional parton distributions, Wigner distributions, etc.
Computing hadronic matrix elements of currents is particularly simple in the light-front, since space-like current matrix elements can be written  as an overlap of frame-independent
LFWFs as in the Drell-Yan-West  formula~\cite{Drell:1969km,West:1970av,Brodsky:1980zm}.   If the virtual space-like photon has $q^+=0$, only processes with the same number of initial and final partons are allowed.    One can also prove fundamental theorems for relativistic quantum field theories using the front form, including: the cluster decomposition theorem~\cite{Brodsky:1985gs}, and the vanishing of the anomalous gravitomagnetic moment for any Fock state of a hadron~\cite{Brodsky:2000ii}.  One also can show that a nonzero anomalous magnetic moment of a bound state requires nonzero angular momentum of the constituents~\cite{Brodsky:1980zm}.
The gauge-invariant meson and baryon distribution amplitudes $\phi_H(x_i)$ which control hard exclusive and direct reactions are the valence LFWFs integrated over transverse momentum at fixed $x_i= {k^+ / P^+}$.   The  ERBL evolution of distribution amplitudes and the factorization theorems for hard exclusive processes  
are derived using LF theory~\cite{Lepage:1980fj,Efremov:1979qk}.  

Making a measurement of a hadron such as DIS $e p \to e^\prime X $  is analogous to taking a flash photograph.   The resulting photograph shows the object as it is exposed at a fixed LF time $\tau$ along the front of the light-wave, not at one instant.  
In fact, the resulting photograph at fixed $\tau$ does not depend on the Lorentz frame of the photographer; i.e.,  whether or not the camera emitting the flash is moving toward the object.  The exposure along the light-front is unchanged.  Analogously, the LFWF measured in DIS is boost invariant. 

Note that a photograph of an extended object at  a fixed ``instant" time $x^0=t$  would require an Avogadro number of simultaneous flashes.  The boost of a wavefunction of a hadron at fixed `instant' time 
 $x^0$  is a difficult nonperturbative dynamical problem~ \cite{Brodsky:1968ea}. Even the particle number changes with the boost.  
 Moreover, the calculation of form factors at fixed instant time 
 $x^0$  requires computing off-diagonal matrix elements, as well as the contributions of currents arising from fluctuations of the vacuum in the initial state which connect to the hadron wavefunction in the final state,  in order to obtain the correct Lorentz invariant result. Thus the knowledge of the wave functions of hadronic eigenstates alone is not sufficient to compute covariant current matrix elements in the usual instant-form framework.  

Light-front time-ordered perturbation theory is equivalent to covariant Feynman perturbation theory.    Cruz-Santiago and Stasto~\cite{Cruz-Santiago:2013vta} have shown that the cluster properties~\cite{Antonuccio:1997tw} of LF time-ordered perturbation theory, together with $J^z$ conservation, can also be used to elegantly derive the Parke-Taylor rules for multi-gluon scattering amplitudes. 

In principle, one can solve nonperturbative QCD by diagonalizing the light-front QCD Hamiltonian $H_{LF}$ directly using the ``discretized light-cone quantization" (DLCQ) method~\cite{Brodsky:1997de} which imposes periodic boundary conditions to discretize the $k^+$ and $k_\perp$ momenta, or the Hamiltonian transverse lattice formulation
introduced in Refs.~\cite{Bardeen:1976tm,Burkardt:2001mf,Bratt:2004wq}. The hadronic spectra and 
LFWFs are then obtained from the eigenvalues and eigenfunctions of the Heisenberg problem $H_{LF} \vert \psi \rangle = M^2 \vert \psi \rangle$, an infinite set of coupled integral equations for the LF components $\psi_n = \langle n \vert \psi \rangle$ in a Fock expansion~\cite{Brodsky:1997de}. 
The DLCQ method has been applied successfully in lower space-time dimensions~\cite{Brodsky:1997de}, such as $QCD(1+1)$~\cite{Pauli:1985pv}. 
For example, one can compute the complete spectrum of meson and baryon states in $QCD(1+1) $ and their LF wavefunctions,  for any number of colors, flavors, and quark masses by matrix diagonalization ~\cite{Hornbostel:1988fb}.  It has also been applied successfully to a range of 1+1 string theory problems by Hellerman and Polchinski~\cite{Hellerman:1999nr,Polchinski:1999br}.  Unlike lattice gauge theory, the DLCQ method is relativistic, has no fermion-doubling,  is formulated in Minkowski space, and is independent of the hadron's momentum $P^+$ and $P_\perp$.
One of the most promising methods for solving nonperturbative (3+1)  QCD is the ``Basis Light-Front Quantization" (BFLQ) method~\cite{Vary:2009gt}. In the BLFQ method one constructs a complete orthonormal basis of eigenstates  based on the eigensolutions of the  effective light-front Schr\"odinger equation derived from light-front holography, in the spirit of the nuclear shell model.  Matrix diagonalization for BLFQ should converge more rapidly than DLCQ since the basis states have a mass spectrum  close to the observed hadronic spectrum.

The light-front vacuum state -- the  eigenstate of  lowest invariant mass $ M$ is trivial up to $k^+=0$ LF zero modes. Thus the LF vacuum is essentially trivial -- there are no quark or gluon vacuum expectation values. The simple structure of the LF vacuum thus allows an unambiguous
definition of the partonic content of a hadron in QCD.   In the case of electroweak theory, the phenomenology is unchanged, except that the Higgs vacuum expectation value of the usual instant-form vacuum is replaced with a light-front zero mode~\cite{Srivastava:2002mw}.  Since the LF vacuum is causal, it can be identified with the observed void of the universe~\cite{Brodsky:2009zd}. 
This has profound implications for the cosmological constant -- the standard contributions from quantum field theory do not contribute~\cite{Brodsky:2012ku,Brodsky:2010xf}.      Thus it is natural in the front form to obtain zero cosmological constant from quantum field theory.

A remarkable feature of the front-form is that five-dimensional anti-de Sitter space (AdS$_5$)
 is holographically dual to light-front Hamiltonian theory for three space dimensions $x_\perp, x^-$  at fixed  LF time $x^+$~\cite{Brodsky:2006uqa, deTeramond:2008ht}.  
 ``Light-front holography" --  the duality between the front form  in physical $3+1$ space-time and 
classical gravity based on the isometries of  AdS$_5$ space --  provides a new method for  determining the eigenstates of the QCD LF Hamiltonian
in the strongly coupled regime.
For example, the valence Fock-state wavefunctions of $H_{LF}$ for zero quark mass obtained from light-front holography satisfy a single-variable relativistic equation of motion in the invariant variable $\zeta^2=b^2_\perp x(1-x)$, which is conjugate to the invariant mass squared ${M^2_{q \bar q} }$. The effective confining potential $U(\zeta^2)$  in this frame-independent ``light-front Schr\"odinger equation" systematically incorporates the effects of higher quark and gluon Fock states~\cite{Brodsky:2013ar}.  The hadron mass scale -- its ``mass gap"  --  is generated in a novel way.
Remarkably,  the potential  $U(\zeta^2)$   has a unique form of a harmonic oscillator potential if one requires that the effective action which underlies the  QCD Lagrangian in the limit of zero quark masses
remains conformally invariant and extends the formalism of de Alfaro, Fubini and Furlan~\cite{deAlfaro:1976je} to light front Hamiltonian theory~\cite{Brodsky:2013ar}.   The result is a nonperturbative relativistic light-front quantum mechanical wave equation which incorporates color confinement and other essential spectroscopic and dynamical features of hadron physics.  A review of light-front holographic methods is given in Ref. \cite{Brodsky:2014yha}.

Finally, we will also show how the physical mass scale $\kappa$ underlying confinement and hadron masses  determines the scale  $\Lambda_s$ in the QCD running coupling.  The relation between scales is obtained by matching the nonperturbative dynamics, as described by  the effective conformal theory mapped to the light-front and its embedding in AdS space, to the perturbative QCD  regime  computed to four-loop order. The result is an effective coupling  defined at all momenta~\cite{Deur:2014qfa}.

\section{The Light-Front Schr\"odinger Equation \label{LFQCD}}

It is advantageous to reduce the full multiparticle eigenvalue problem of the LF Hamiltonian to an effective LF Schr\"odinger equation  which acts on the valence sector LF wavefunction and determines each eigensolution separately~\cite{Pauli:1998tf}.   In contrast,  diagonalizing the LF Hamiltonian yields all eigensolutions simultaneously, a complex task. The central problem  then becomes the derivation of the effective interaction $U$ which acts only on the valence sector of the theory and has, by definition, the same eigenvalue spectrum as the initial Hamiltonian problem.  In order to carry out this program one must systematically express the higher Fock components as functionals of the lower ones. This method has the advantage that the Fock space is not truncated, and the symmetries of the Lagrangian are preserved~\cite{Pauli:1998tf}.   

To a first approximation, LF QCD is formally equivalent to the equations of motion on a fixed gravitational background asymptotic to AdS$_5$~\cite{deTeramond:2008ht}.   In fact, the usual introduction of a dilaton profile is equivalent to a modification of the AdS metric but it is left largely unspecified. However, we shall show, if one imposes the requirement that  the action of the corresponding one-dimensional effective theory  remains conformally invariant, then the dilaton profile  is constrained to be quadratic,  a remarkable result which follows from the dAFF construction of conformally invariant quantum mechanics~\cite{Brodsky:2013ar}.   A related argument is given in Ref.~\cite{Glazek:2013jba}.

A hadron has four-momentum $P = (P^-, P^+,  \mbf{P}_\perp)$, $P^\pm = P^0 \pm P^3$ and invariant mass $P^2 = M^2$. The generators $P = (P^-, P^+,  \vec{P}_\perp)$ are constructed canonically from the QCD Lagrangian by quantizing the system on the light-front at fixed LF time $x^+$, $x^\pm = x^0 \pm x^3$~\cite{Brodsky:1997de}. 
The LF Hamiltonian $P^-$ generates the LF time evolution with respect to the LF time  $x^+$, whereas the LF longitudinal $P^+$ and transverse momentum $\vec P_\perp$ are kinematical generators.  
In the limit of zero quark masses the longitudinal modes decouple  from the  invariant  LF Hamiltonian  equation  $H_{LF} \vert \phi \rangle  =  M^2 \vert \phi \rangle$,  with  $H_{LF} = P_\mu P^\mu  =  P^- P^+ -  \mbf{P}_\perp^2$.  The result is a relativistic and frame-independent  wave equation for $\phi$~\cite{deTeramond:2008ht} 
\begin{equation} \label{LFWE}
\left[-\frac{d^2}{d\zeta^2}
- \frac{1 - 4L^2}{4\zeta^2} + U\left(\zeta^2, J\right) \right]
\phi_{n,J,L}(\zeta^2) = 
M^2 \phi_{n,J,L}(\zeta^2).
\end{equation}
The effective interaction $U(\zeta^2,J)$ is instantaneous in LF time and acts on the lowest state of the LF Hamiltonian. This equation describes the spectrum of mesons as a function of $n$, the number of nodes in $\zeta$, the total angular momentum  $J$, which represents the maximum value of $\vert J^z \vert$, $J = \max \vert J^z \vert$,
and the internal orbital angular momentum of the constituents $L= \max \vert L^z\vert$. The  ${\rm SO(2)}$ Casimir  $L^2$  corresponds to the group of rotations in the transverse LF plane. By using the dictionary of light-front holography, the variable $z$ of AdS space becomes identified with the LF   boost-invariant transverse-impact variable $\zeta$~\cite{Brodsky:2006uqa}, 
thus giving the holographic variable a precise definition in LF QCD~\cite{Brodsky:2006uqa, deTeramond:2008ht}.
For a two-parton bound state $\zeta^2 = x(1-x) b^{\,2}_\perp$.

In the exact QCD theory the potential in the LF Schr\"odinger equation (\ref{LFWE}) is determined from the two-particle irreducible (2PI) $ q \bar q \to q \bar q $ Greens' function.  In particular, the reduction from higher Fock states in the intermediate states
leads to an effective interaction $U\left(\zeta^2, J\right)$  for the valence $\vert q \bar q \rangle$ Fock state~\cite{Pauli:1998tf}.
The LF wave equation is the relativistic frame-independent front-form analog of the non-relativistic radial Schr\"odinger equation for muonium  and other hydrogenic atoms in presence of an instantaneous Coulomb potential;  it could emerge from the exact QCD formulation when one includes in the confinement potential contributions which are due to the exchange of two connected gluons; {\it i.e.}, ``H'' diagrams~\cite{Appelquist:1977tw}.
We notice that $U$ becomes complex for an excited state since a denominator can vanish; this gives a complex eigenvalue and the decay width.  The multi-gluon exchange diagrams also could be connected to the Isgur-Paton~\cite{Isgur:1984bm} flux-tube model of confinement; the collision of flux tubes could give rise to the ridge phenomena recently observed in high energy $pp$ collisions at RHIC~\cite{Bjorken:2013boa}. A related approach for determining the valence LF wavefunction and studying the effects of higher Fock states without truncation has been given in Ref.~\cite{Chabysheva:2011ed}.   Unlike ordinary instant-time quantization, the LF Hamiltonian equations of motion are frame independent; remarkably, they  have a structure which matches exactly the eigenmode equations in AdS space~\cite{deTeramond:2008ht}. This makes a direct connection of QCD with AdS methods possible.     One also can show that the linear potential used as a leading approximation for describing color confinement of heavy quarks in the instant form of dynamics corresponds to a harmonic oscillator confining potential in the front form of dynamics~\cite{Trawinski:2014msa}.

\section{Effective Confinement from the Gauge/Gravity Correspondence}

The  correspondence between AdS and LF QCD was originally motivated~\cite{Brodsky:2006uqa} by the AdS/CFT correspondence between gravity on a higher-dimensional space and conformal field theories in physical space-time~\cite{Maldacena:1997re}.  It has as its most explicit and simplest realization  a direct holographic mapping to LF Hamiltonian theory~\cite{deTeramond:2008ht}, which provides a precise relation between the wavefunctions in AdS space and the boost-invariant bound-state LF wavefunctions describing the internal structure of hadrons in physical space-time.  
The resulting valence Fock-state wavefunctions satisfy a single-variable relativistic equation of motion analogous to the eigensolutions of the nonrelativistic radial Schr\"odinger equation.  There is a precise  connection between the quantities that enter the fifth dimensional AdS space and the physical variables of LF theory.  For example, the AdS mass  parameter $\mu R$ maps to the LF orbital angular momentum.  The formulae for electromagnetic~\cite{Polchinski:2002jw} and gravitational~\cite{Abidin:2008ku} form factors in AdS space map to the exact Drell-Yan-West formulae in LF QCD~\cite{Brodsky:2006uqa, Brodsky:2007hb, Brodsky:2008pf}.

Recently we have derived wave equations for hadrons with arbitrary spin $J$ starting from an effective action in  AdS space~\cite{deTeramond:2013it}.    An essential element is the mapping of the higher-dimensional equations  to the LF Hamiltonian equation  found in Ref.~\cite {deTeramond:2008ht}.  This procedure allows a clear distinction between the kinematical and dynamical aspects of the LF holographic approach to hadron physics.  Accordingly, the non-trivial geometry of pure AdS space encodes the kinematics,  and the additional deformations of AdS encode the dynamics, including confinement~\cite{deTeramond:2013it}.  Upon the substitution of the holographic variable $z$ by the LF invariant variable $\zeta$ and replacing $\Phi_J(z)   = \left(R/z\right)^{J- (d-1)/2} e^{-\varphi(z)/2} \, \phi_J(z)$ 
in the AdS wave equation for general spin $J$,  we find from the dilaton-modified AdS action the effective LF potential~\cite{deTeramond:2013it, deTeramond:2010ge}
\begin{equation} \label{U}
U(\zeta^2, J) = \frac{1}{2}\varphi''(\zeta^2) +\frac{1}{4} \varphi'(\zeta^2)^2  + \frac{2J - 3}{2 \zeta} \varphi'(\zeta^2) ,
\end{equation}
provided that the AdS mass $\mu$ is related to the internal orbital angular momentum $L = max \vert L^z \vert$ and the total angular momentum $J^z = L^z + S^z$ according to $(\mu \, R)^2 = - (2-J)^2 + L^2$.  The critical value  $L=0$  corresponds to the lowest possible stable solution, the ground state of the LF Hamiltonian.
For $J = 0$ the five dimensional mass $\mu$
 is related to the orbital  momentum of the hadronic bound state by
 $(\mu \, R)^2 = - 4 + L^2$ and thus  $(\mu\, R)^2 \ge - 4$. The quantum mechanical stability condition $L^2 \ge 0$ is thus equivalent to the Breitenlohner-Freedman stability bound in AdS~\cite{Breitenlohner:1982jf}.

\begin{figure}[h]
\centering
\includegraphics[width=5.8cm]{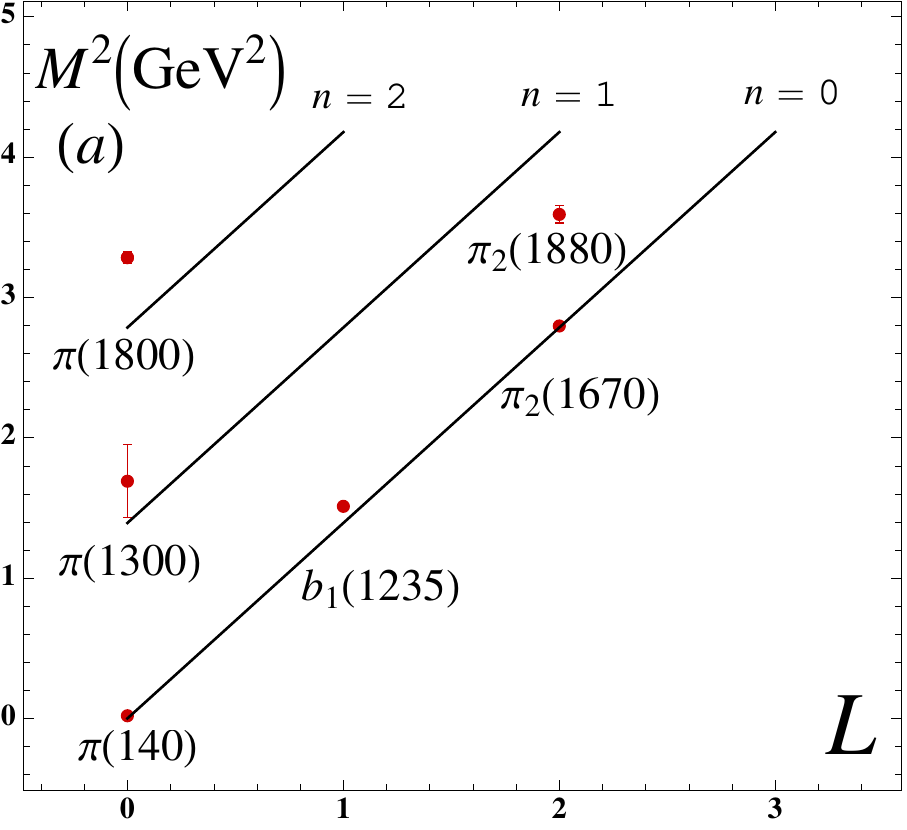}  \hspace{30pt}
\includegraphics[width=5.8cm]{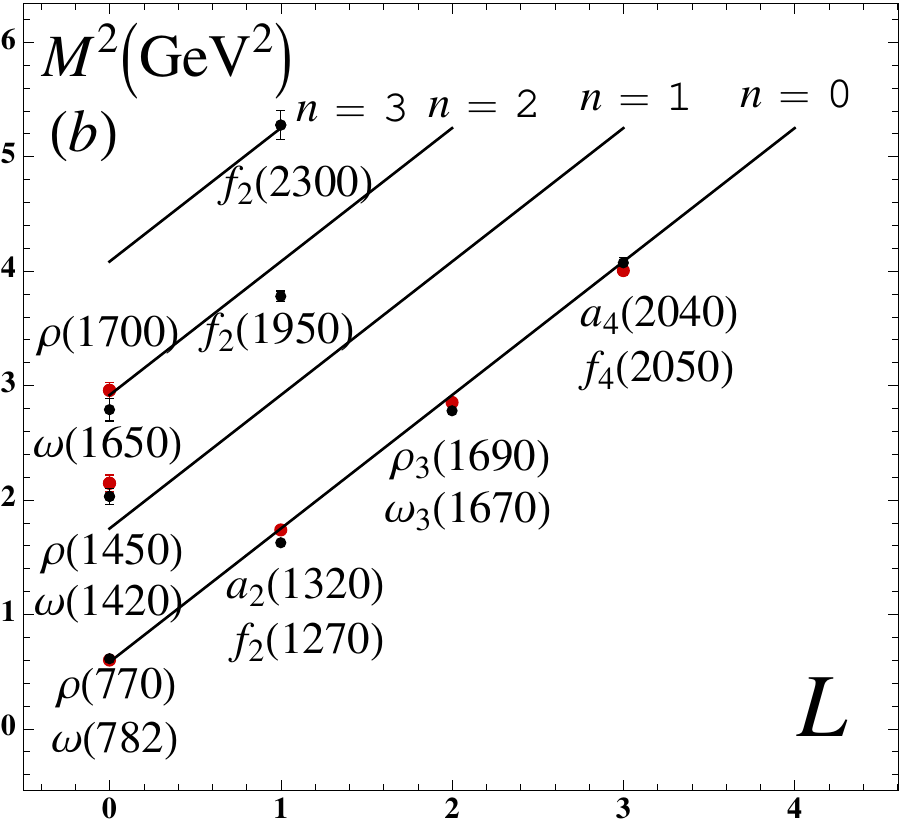}
 \caption{\small $I\!=\!1$ parent and daughter Regge trajectories for the light pseudoscalar mesons (a) with
$\kappa = 0.59$ GeV; and  $I=0$ and $I=1$  light vector-mesons
 (b) with $\kappa= 0.54$ GeV.}
\label{pionspec}
\end{figure} 

The correspondence between the LF and AdS equations  thus determines the effective confining interaction $U$ in terms of the infrared behavior of AdS space and gives the holographic variable $z$ a kinematical interpretation. 
The dilaton profile $\exp{\left(\pm \kappa^2 z^2\right)}$  leads immediately to linear Regge trajectories~\cite{Karch:2006pv}.
For the  confining solution $\varphi = \exp{\left(\kappa^2 z^2\right)}$, the effective potential is
$U(\zeta^2,J) =   \kappa^4 \zeta^2 + 2 \kappa^2(J - 1)$ and  leads to eigenvalues
$M_{n, J, L}^2 = 4 \kappa^2 \left(n + \frac{J+L}{2} \right)$,
with a string Regge form $M^2 \sim n + L$.  
A detailed discussion of the light meson and baryon spectrum,  as well as  the elastic and transition form factors of the light hadrons using LF holographic methods, is given in 
Refs. \cite{Brodsky:2014yha} and \cite{Brodsky:2014yha}. As an example, the spectral predictions  for the $J = L + S$ light pseudoscalar and vector meson  states are  compared with experimental data in Fig. \ref{pionspec} for the positive sign dilaton model.

\section{Uniqueness of the Confining Potential}

If one sets the masses of the quarks to zero, no mass scale appears explicitly in the QCD Lagrangian, displaying invariance under both scale  (dilatation) and special conformal transformations~\cite{Parisi:1972zy}.   Nevertheless, the theory built upon this conformal template displays color confinement,  a mass gap, as well as asymptotic freedom. 
A  fundamental question is thus how does the mass scale which determines the masses of the light-quark hadrons,  the range of color confinement, and the running of the  coupling emerge in QCD? 
The effective  confinement potential which appears in the LF equations of motion is unique if one requires that the corresponding one-dimensional effective action which encodes the chiral symmetry of QCD remains conformally invariant. In addition, Leutwyler and Stern~\cite{Leutwyler:1977vy}  have also argued that the only possible potential that can appear in the semiclassical light-front equation of motion that acts on the valence Fock state in a transverse variable must also take the form of a harmonic oscillator.

We start with the one-dimensional action~\cite{deAlfaro:1976je}.
$
{ \cal{S}}= \half \int dt \left (\dot Q^2 - \frac{g}{Q^2} \right),
$
which  is invariant under conformal transformations in the 
variable $t$.   The scale-invariant $1/Q^2$ term is the analog of a centrifugal term in the kinetic energy. In addition to the Hamiltonian  $H(Q, \dot Q) = \half \Big(\dot Q^2 + \frac{g}{Q ^2}\Big)$ there are two more invariants of motion for this field theory, namely the 
dilation operator $D$ and the  special conformal transformation operator $K.$
Specifically, if one introduces the new variable $\tau$ defined through 
$d\tau= d t/(u+v\,t + w\,t^2)$ and the  rescaled fields $q(\tau) = Q(t)/(u + v\, t + w \,t^2)^{1/2}$,
it then follows that the  operator
$G= u\,H_t + v\,D + w\,K$ which generates the quantum mechanical 
evolution in  $\tau$~\cite{deAlfaro:1976je} is a compact operator and thus  introduces a mass scale.
The new Hamiltonian  operator $G$ is a linear combination of the old Hamiltonian $H$, $D$ and $K$, where $u$, $v$ and $w$ are arbitrary coefficients. 
One can show explicitly~\cite{deAlfaro:1976je, Brodsky:2013ar} that  a confinement length scale appears in the action when one expresses the action  in terms of  the new time variable $\tau$ and the new fields $q(\tau)$, without affecting its conformal invariance. Furthermore, for $g \geq -1/4$ and $4\, u w -v^2 > 0$ the 
corresponding Hamiltonian $G(q, \dot q)=  \half \Big( \dot q^2 + \frac{g}{q^2} +\frac{4\,uw - v^2}{4} q^2\Big)$ is a compact operator.  Finally, we can transform back to the original field operator $Q(t)$.  We find 
\beqa \label{HtauQ}
G(Q, \dot Q ) \! &= \! & \frac{1}{2} u \Big(\dot Q^2 + \frac{g}{Q^2} \Big)  - \frac{1}{4} v \Big( Q \dot Q + \dot Q Q\Big) + \frac{1}{2} w Q^2\\  \nn
            \! &= \! & u H_t + v D + w K,
            \enqa
at $t=0$.  We thus recover the evolution operator $G =  u H_t + v D + w K$ which describes the evolution in the variable $\tau$, but expressed in terms of the original field $Q$.
The Shr\"odinger picture follows by identifying $Q \to x$ and $\dot Q \to -i {d\over dx}$. 
Then the evolution operator in the new time variable $\tau$ is
\beq \label{Htaux} 
G = {1\over 2} u \Big(-{d^2\over dx^2}  + {g\over x^2} \Big) + {i\over 4} v \Big(x {d\over dx} + {d\over dx}x \Big) +{1\over 2}wx^2
\enq

If we now compare   the Hamiltonian \req{Htaux}  with the LF wave equation \req{LFWE} and identify the variable $x$ with the LF invariant variable $\zeta$,  we can identify $u=2, \; v=0$ and relate the dimensionless constant $g$ to the LF orbital angular momentum, $g=L^2-1/4$,  in order to reproduce the LF kinematics. Furthermore  $w$ fixes the confining LF  potential to a quadratic $\la^2 \, \zeta^2$ dependence.
The mass scale brought in via $w = 2\kappa^2$ then generates the confining mass scale $\kappa$.
The dilaton is also unique: $e^{\phi(z)} = e^{\pm \kappa^2 z^2},$ where  $z^2$ is matched to $\zeta^2=b^2_\perp x(1-x)$ via LF holography. 
The spin-$J$ representations in AdS$_5$~\cite{deTeramond:2013it}  then lead uniquely to the LF confining potential 
$U(\zeta^2) = \kappa^4 \zeta^2 - 2 \kappa^2(J-1)$, since only the positive sign dilaton profile is compatible with LF holographic mapping~\cite{deTeramond:2013it}.
The new time variable $\tau$ is related to the variable $t$ for the case  $u w >0, \,v=0$   by 
$\tau =\frac{1}{\sqrt{u\,w}} \arctan\left(\sqrt{\frac{w}{u}} t\right)$,
{\it i.e.} $\tau$ has only a limited range, characteristic of a relativistic bound state.
The finite range of invariant LF time $\tau=x^+/P^+$ can be interpreted as a feature of the internal frame-independent LF  time difference between the confined constituents in a bound state. For example, in the collision of two mesons, it would allow one to compute the LF time difference between the two possible quark-quark collisions.
The QCD mass scale $\kappa$ in units of GeV has to be determined by one measurement; e.g., the proton mass. All other masses and size parameters are then predicted.

We have applied  LF holography to describe the full baryon spectrum~\cite{deTeramond:2014yga}, the $K$-meson spectrum (including quark masses)~\cite{deTeramond:2014rsa}, space-like and time-like form factors~\cite{deTeramond:2012rt}, as well as transition amplitudes such as 
$\gamma^* \gamma \to \pi^0$~\cite{Brodsky:2011xx}, $\gamma^* N \to N^*$~\cite{deTeramond:2011qp}, all based on essentially the single  mass scale parameter $\kappa.$  Many other applications have been presented in the literature, including recent results by Forshaw and Sandapen~\cite{Forshaw:2012im} for diffractive  $\rho$ electroproduction which are  based on the LF holographic prediction for the longitudinal $\rho$ LFWF.  Other recent applications include predictions for  generalized parton distributions (GPDs)~\cite{Vega:2010ns}, and a model for nucleon and flavor form factors~\cite{Chakrabarti:2013dda}. 

The separate dependence on $J$ and $L$ in the meson spectrum leads to a  mass ratio of the $\rho$ and the $a_1$ mesons which coincides with the result of the Weinberg sum rules~\cite{Weinberg:1967kj}. 
The treatment of the chiral limit in the LF holographic approach to strongly coupled QCD is substantially different from the standard approach  based on chiral perturbation theory.
In the color-confining 
LF holographic model discussed here, the vanishing of the pion mass in the chiral limit, a phenomenon usually ascribed to spontaneous symmetry breaking of the chiral symmetry,  is  obtained specifically from the precise cancellation of the LF kinetic energy and LF potential energy terms for the quadratic confinement potential. This mechanism provides a  viable alternative to the conventional description of nonperturbative QCD based on vacuum condensates~\cite{Shifman:1978bx}, and it 
eliminates a major conflict of hadron physics with the empirical value for the cosmological constant~\cite{Brodsky:2009zd,Brodsky:2010xf}.

\section{Scheme-Independent Determination of the Perturbative QCD Scale $\bf \Lambda_s$ from Confinement Dynamics in Holographic QCD}

One of the fundamental questions in QCD is the relationship between the hadronic mass scale and the mass parameter $\Lambda_{\overline{MS}}$, which controls the evolution of the running coupling $\alpha_{\overline {MS}}(Q^2)$.   Since the perturbative coupling apparently diverges at $Q^2 = \Lambda^2_{\overline{MS}}$,  it is often argued, based on dimensional transmutation,  that the confinement scale  $\kappa$, could in principle, be computed starting from the empirically determined value of $\Lambda_{\overline {MS}}$.  A serious complication which immediately confronts this procedure is the fact that  $\Lambda_{\overline {MS}}$ depends on the choice of the ${\overline {MS}}$ renormalization scheme, whereas the value of $\kappa$ and hadron masses are scheme-independent.  
The logic of dimensional transmutation can in fact be reversed:  the mass scale $\kappa$ underlying color confinement and hadron masses can be taken as the fundamental parameter of QCD.  We then use an analytic method which connects the nonpertubative and perturbative domains of QCD to determine the scale parameter $\Lambda_s$. The method can be applied to any renormalization scheme  (RS),  including  the $ {\overline {MS}}$ scheme~\cite{Deur:2014qfa}.  We will demonstrate this connection specifically in the effective theory based on holographic QCD~\cite{Brodsky:2006uqa, deTeramond:2008ht, Brodsky:2013ar}. 
As we have discussed, the LF holographic approach to the strong interactions originates from a remarkable connection of LF dynamics to classical gravity in a higher dimensional  Anti-de Sitter (AdS) space, as well as the properties of conformally invariant one-dimensional effective quantum field~\cite{deAlfaro:1976je}.  This connects the form of the effective LF confining interaction holographically to a unique dilaton profile in the embedding AdS space~\cite{Brodsky:2013ar}. 

\begin{figure}
\centering
\includegraphics[width=6.5cm]{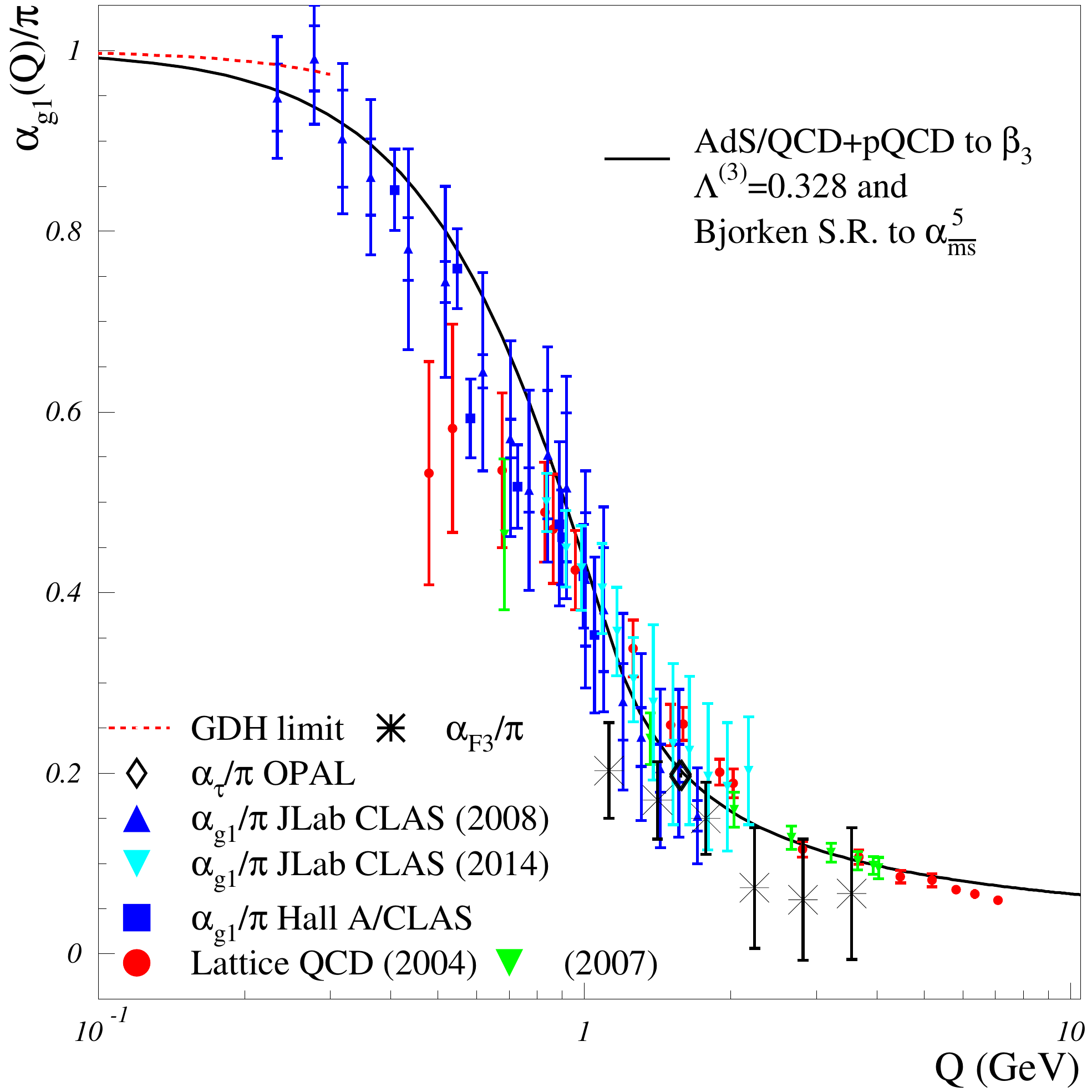}
\caption{\label{Fig:comparison with data} Comparison between the analytical expression for the effective charge $\alpha_{g_{1}}$ obtained from  the matching procedure of  hard and soft regimes  with the experimental JLab data, lattice QCD results, and the GDH sum rule constraint.}
\end{figure}

As Grunberg~\cite{Grunberg:1980ja}  has emphasized, it is natural to define the QCD coupling from a physical observable which is perturbatively calculable at large $Q^2$. This is analogous to QED,  where the standard running Gell Mann-Low coupling $\alpha(t)$ is defined from the elastic scattering amplitude of heavy leptons.  A physically defined  ``effective charge"  in QCD incorporates nonperturbative dynamics at low scales, and then evolves to the familiar pQCD form $4\pi / \beta_0 \log\left(Q^2/\Lambda_s^2\right)$ as required by asymptotic freedom at high scales.   As expected on physical grounds effective charges are  infrared (IR) finite and smooth at small virtualities~\cite{Cvetic:2013gta}. Examples of freezing of the coupling at low momenta (IR fixed point) are described in~\cite{Cornwall:1981zr, Shirkov:1997wi, Fischer:2003rp, Aguilar:2009nf, Courtoy:2013apa} (a full reference list is given in~\cite{Cvetic:2013gta}).  
We shall focus  on the effective coupling $\alpha_{g_1} = g^2_1/4 \pi$  defined from the  Bjorken sum rule~\cite{Bjorken:1966jh}, which is the best-measured effective charge~\cite{Deur:2005cf,Deur:2008rf}:
\begin{equation} \label{BjSR}
\frac{\alpha_{g_{1}}(Q^2)}{\pi} = 1 -  \frac{6}{g_{A}} \int_{0}^{1} dx \, g_{1}^{p-n}(x,Q^{2}),
\end{equation}
where $x$ is the Bjorken scaling variable, $g_1^{p-n}$  is the isovector component of the nucleon first spin structure function and $g_A$ is the nucleon axial charge. The effective charge $\alpha_{g_1}(Q^2)$ is kinematically constrained to satisfy $\alpha_{g_1}\left(Q^2=0\right)=\pi$, and it is thus IR finite. The Gerasimov-Drell-Hearn (GDH)  sum rule~\cite{Drell:1966jv, Gerasimov:1965et}  implies that $\alpha_{g_1}(Q^2)$ is nearly conformal in the IR domain~\cite{Deur:2005cf}. 
The coupling $\alpha_{g_1}(Q^2)$ can be extrapolated to large distances.  In the low $Q^2 < 1 $  GeV$^2$ domain the Jefferson Lab data~\cite{Deur:2005cf} for  $\alpha_{g_1}(Q^2)$  are remarkably consistent with the  Gaussian form  predicted by LF holographic QCD~\cite{Brodsky:2010ur}.
\begin{equation} \label{alphaAdS}
\alpha^{AdS}_{g_1}(Q^2) = \pi \exp{\left(-Q^2/4 \kappa^2\right)},
\end{equation}
At high $Q^2$ one can compute $\alpha_{g_1}(Q^2)$ in pQCD as a perturbative expansion in $\alpha_{\overline{MS}}(Q^2)$ which is presently known up to five loops.  The complete functional dependence of $\alpha_{g_1}(Q^2)$  must be analytic.  We thus expect an analytically smooth transition between the soft and hard pQCD regimes at a scale $Q_0 \simeq 2 \kappa \simeq 1$ GeV. Thus, we can approximate the transition between the soft and hard domains  by matching the holographic and pQCD forms, for example by  imposing continuity of the couplings and its first derivative which fixes  the transition scale $Q_0$. We then can  use the pQCD prediction for $\alpha_{g_1}(Q^2)$ in powers of $\alpha_{\overline{MS}}(Q^2)$ to obtain a value of $\Lambda_{\overline {MS}}$ from  the scheme-independent scale $\kappa$.  Commensurate scale relations between couplings allow this procedure to be applied to any choice of renormalization scheme~\cite{Brodsky:1994eh, Brodsky:1995tb, Wu:2013ei}.
Since $Q_0$ is relatively small, higher orders in perturbation theory will be essential for an accurate determination of $\Lambda_s$  and to evaluate the convergence of the result and its uncertainty. In addition, since we determine $\Lambda_s$ by imposing continuity conditions at a transition point $Q_0$ -- which is in turn determined by continuity conditions -- it is important that there is a rapid, smooth transition between the two regimes~\cite{Deur:2004ti,Deur:2014vea,Courtoy:2013apa}. The parton-hadron duality 
\cite{Bloom:1970xb}  provides a general argument why the transition between perturbative and nonperturbative dynamics occurs over a relatively small domain of $Q^2$. 
The pQCD $\beta$-series for $\alpha_{g1}$ has not been computed. However, $\alpha_{\overline{MS}}$ has been calculated up to 4 loops (up to $\beta_{3}$). One can then transform from the $\overline{MS}$ to the $g_{1}$ scheme by using the perturbative $\overline{MS}$ expression of the Bjorken sum rule~\cite{Kataev:1994gd}
\begin{multline}  \label{eq: alpha_g1 from Bj SR}
\alpha_{g_{1}}^{pQCD}(Q^{2})  = \pi \Big[\frac{\alpha_{\overline{MS}}}{\pi} + 3.58\left(\frac{\alpha_{\overline{MS}}}{\pi}\right)^{2} + 20.21\left(\frac{\alpha_{\overline{MS}}}{\pi}\right)^{3} + 130\left(\frac{\alpha_{\overline{MS}}}{\pi}\right)^{4} + 893.38\left(\frac{\alpha_{\overline{MS}}}{\pi}\right)^{5} + \cdots \Big].
\end{multline}
At a fixed $Q^{2}$, the equality between $\alpha_{g_{1}}^{pQCD}$, Eq. (\ref{eq: alpha_g1 from Bj SR}), and $\alpha_{g_{1}}^{AdS}$, Eq.  (\ref{alphaAdS}), is fulfilled for a unique value of $\Lambda_{\overline{MS}}$ since, at fixed $Q^{2}$, $\alpha_{s}$ is monotonic with $\Lambda_{\overline{MS}}$. 
Eq. (\ref{eq: alpha_g1 from Bj SR}) is an expansion in $\alpha_{\overline{MS}}$, itself a series in  $\beta_{n}$. The  value of $\Lambda_{\overline{MS}}$ and the shape of $\alpha_{g_{1}}(Q^{2})$ depend on the orders to which these two series are truncated. To compare consistently to the world data the $\beta$ series is  truncated at $\beta_{3}$, while to  provide a transformation from the $\overline{MS}$ to the $g_{1}$ schemes as complete as possible, the order of the $\alpha_{\overline{MS}}$
series is  kept as high as possible. Remarkably, the curves converge quickly to an universal shape essentially invariant with the $\beta_{n}$ and $\alpha_{\overline{MS}}$ orders:    at orders  $\beta_{n}$ or $\alpha_{\overline{MS}}^{n}$, $n > 1$, the  couplings $\alpha_{g_{1}}(Q^{2})$ are nearly identical. 
The value of $\Lambda_{\overline{MS}}$, however,  does depend significantly on the truncation order of Eq. (\ref{eq: alpha_g1 from Bj SR}). 
However,  the results are consistent  at each order with the PDG 2012 data.
Our result at highest order is $\Lambda^{(3)}_{\overline{MS}} = 0.328 \pm 0.034$ GeV at $\beta_{3}$ and for $n_{f} = 3$~\cite{Deur:2014qfa},   where the uncertainty is  derived from the truncation of the series  for $\alpha_{g_1}$  and a theoretical uncertainty of $\pm \, 0.024$ GeV from the extraction of $\kappa$  from the $\rho$ or the  proton mass. We have neglected the uncertainty from  the chiral limit  extraction of $\kappa$ which amounts to $\pm \, 0.003$ GeV. Our result is to be compared with  the world data $\Lambda_{\overline{MS}}^{(3)} = 0.339 \pm 0.010$ GeV.   In Fig. \ref{Fig:comparison with data} we compare  our prediction with the experimental and lattice results for $\alpha_{g1}$.

The strong coupling constant $\alpha_s$ is also computable in Lattice QCD (LQCD)~\cite{Lepage:1991zt, Davies:2003ik}. Although there is a parallel between the LQCD procedure and our matching procedure, notable differences are that LF holographic QCD has only one parameter, the confinement scale  $\kappa$, whereas  in addition to the bare coupling  LQCD has four other parameters~\cite{Davies:2003ik}.  
On the other hand, the foundation of LQCD stems directly from the QCD Lagrangian while LF holographic QCD is an effective theory. 
These differences underline the complementarity of the two approaches. LQCD provides by far the best determination of $\Lambda_{\overline{MS}}$, or equivalently of $\alpha_{\overline{MS}}$, with a relative uncertainty of $0.6\%$. This can be compared with $1.3\%$ and $1.8\%$, the uncertainties on the two best experimental results, respectively from the world data on $\tau$-decays and DIS. Our uncertainty is $1.7\%$, similar to the accuracy of the combined world DIS data, and would be similar to the $\tau$ world data if Eq. (\ref{eq: alpha_g1 from Bj SR}) was available to the next order.

\section{Summary}

One of the most profound questions in QCD is the origin of the mass scale which determines the range of color confinement and hadron masses.  As shown by dAFF, a mass scale $\kappa$ can appear in the Hamiltonian and the equations of motion without affecting the conformal invariance of the action~\cite{deAlfaro:1976je}.  When one applies the dAFF procedure to LF Hamiltonian theory, the result is a unique effective confining  potential and a unique  dilaton profile in the embedding AdS  space~\cite{Brodsky:2013ar} which determines the higher-spin representations~\cite{deTeramond:2013it}. The eigenvalues of the resulting frame-independent light-front Hamiltonian accurately describe the light-quark spectroscopy of hadrons in terms of the fundamental mass scale $\kappa$ ({\it e.g.} $M_\rho = \sqrt 2 \, \kappa$ and $M_\pi=0$ for massless quarks), as well as dynamical observables such as hadronic form factors~\cite{Brodsky:2014yha} and diffractive vector meson electroproduction~\cite{Forshaw:2012im}.  

We have  shown how the physical  scale $\kappa$ (or a single hadron mass such as $m_\rho$  also determines the scale parameter $\Lambda_s$ controlling the evolution of the QCD running coupling $\alpha_s$.  The relation between these scales is obtained by matching the nonperturbative dynamics of the coupling, as predicted by light-front holography~\cite{Brodsky:2010ur},  to the perturbative dynamics of QCD in the high momentum transfer regime.  This procedure determines the  effective charge $ \alpha_{g1}(Q^2)$ for all $Q^2$. At low values of $Q^2$,  the coupling takes the holographic form (\ref{alphaAdS}), which describes well the measurements of the Bjorken sum rule~\cite{Deur:2005cf}.  At  high $Q^2$ the logarithmic pQCD evolution of $\alpha_{g1}(Q^2)$  is known to five-loops in the $\overline{MS}$ scheme. The matching  of the  $ \alpha_{g1}(Q^2)$  coupling and its slope at the intersection of the perturbative and nonperturbative  domains  determines  $ \Lambda_{\overline{MS}}$ in terms of $\kappa$,  as well as the transition scale $Q_0 \simeq 2 \kappa \simeq1$ GeV. The predicted value $\Lambda_{\overline{MS}} = 0.328 \pm 0.034$ GeV, is in agreement with the world average $0.339 \pm 0.010$ GeV.  The relation between scales is $\Lambda_{\overline {MS}}=  (0.599 \pm 0.062) \, \kappa = (0.423 \pm 0.044) \, M_\rho$.  This analysis can also be applied to other effective charges and other choices of renormalization schemes.

\begin{acknowledgements}
Invited talk, presented by SJB at 
{\it Theory and Experiment for Hadrons on the Light-Front (Light Cone 2014) }
May 26 -- 30, 2013,  Raleigh, North Carolina.  We thank Professor Chueng-Ryong Ji for organizing this outstanding meeting. 
This material is based in part upon work supported by the U.S. Department of Energy, Office of Science, Office of Nuclear Physics under contract DE-AC05-06OR23177
and the U.S.  Department of Energy contract DE--AC02--76SF00515.    
SLAC-PUB-16098.
\end{acknowledgements}


\begin{thebibliography}{3}



\bibitem{Dirac:1949cp} 
  P.~A.~M.~Dirac,
  ``Forms of Relativistic Dynamics,''
  Rev.\ Mod.\ Phys.\  {\bf 21}, 392 (1949).


\bibitem{Brodsky:1997de} 
  S.~J.~Brodsky, H.~C.~Pauli and S.~S.~Pinsky,
  ``Quantum chromodynamics and other field theories on the light cone,''
  Phys.\ Rept.\  {\bf 301}, 299 (1998)
  [hep-ph/9705477].


\bibitem{Drell:1969km} 
  S.~D.~Drell and T.~M.~Yan,
  ``Connection of Elastic Electromagnetic Nucleon Form-Factors at Large Q**2 and Deep Inelastic Structure Functions Near Threshold,''
  Phys.\ Rev.\ Lett.\  {\bf 24}, 181 (1970).


\bibitem{West:1970av} 
  G.~B.~West,
  ``Phenomenological model for the electromagnetic structure of the proton,''
  Phys.\ Rev.\ Lett.\  {\bf 24}, 1206 (1970).


\bibitem{Brodsky:1980zm} 
  S.~J.~Brodsky and S.~D.~Drell,
  ``The Anomalous Magnetic Moment and Limits on Fermion Substructure,''
  Phys.\ Rev.\ D {\bf 22}, 2236 (1980).


\bibitem{Brodsky:1985gs} 
  S.~J.~Brodsky and C.~R.~Ji,
  ``Factorization Property of the Deuteron,''
  Phys.\ Rev.\ D {\bf 33}, 2653 (1986).


\bibitem{Brodsky:2000ii} 
  S.~J.~Brodsky, D.~S.~Hwang, B.~Q.~Ma and I.~Schmidt,
  ``Light cone representation of the spin and orbital angular momentum of relativistic composite systems,''
  Nucl.\ Phys.\ B {\bf 593}, 311 (2001)
  [hep-th/0003082].


\bibitem{Lepage:1980fj} 
  G.~P.~Lepage and S.~J.~Brodsky,
  ``Exclusive Processes in Perturbative Quantum Chromodynamics,''
  Phys.\ Rev.\ D {\bf 22}, 2157 (1980).


\bibitem{Efremov:1979qk} 
  A.~V.~Efremov and A.~V.~Radyushkin,
  ``Factorization and Asymptotical Behavior of Pion Form-Factor in QCD,''
  Phys.\ Lett.\ B {\bf 94}, 245 (1980).


\bibitem{Brodsky:1968ea} 
  S.~J.~Brodsky and J.~R.~Primack,
  ``The Electromagnetic Interactions of Composite Systems,''
  Annals Phys.\  {\bf 52}, 315 (1969).


\bibitem{Cruz-Santiago:2013vta} 
  C.~A.~Cruz-Santiago and A.~M.~Stasto,
  ``Recursion relations and scattering amplitudes in the light-front formalism,''
  Nucl.\ Phys.\ B {\bf 875}, 368 (2013)
  [arXiv:1308.1062 [hep-ph]].


\bibitem{Antonuccio:1997tw} 
  F.~Antonuccio, S.~J.~Brodsky and S.~Dalley,
  ``Light cone wave functions at small x,''
  Phys.\ Lett.\ B {\bf 412}, 104 (1997)
  [hep-ph/9705413].


\bibitem{Bardeen:1976tm} 
  W.~A.~Bardeen and R.~B.~Pearson,
  ``Local Gauge Invariance and the Bound State Nature of Hadrons,''
  Phys.\ Rev.\ D {\bf 14}, 547 (1976).


\bibitem{Burkardt:2001mf} 
  M.~Burkardt and S.~K.~Seal,
  ``A Study of light mesons on the transverse lattice,''
  Phys.\ Rev.\ D {\bf 65}, 034501 (2002)
  [hep-ph/0102245].


\bibitem{Bratt:2004wq} 
  J.~Bratt, S.~Dalley, B.~van de Sande and E.~M.~Watson,
  ``Small-x behaviour of lightcone wavefunctions in transverse lattice gauge theory,''
  Phys.\ Rev.\ D {\bf 70}, 114502 (2004)
  [hep-ph/0410188].


\bibitem{Pauli:1985pv} 
  H.~C.~Pauli and S.~J.~Brodsky,
  ``Solving Field Theory in One Space One Time Dimension,''
  Phys.\ Rev.\ D {\bf 32}, 1993 (1985).


\bibitem{Hornbostel:1988fb} 
  K.~Hornbostel, S.~J.~Brodsky and H.~C.~Pauli,
  ``Light Cone Quantized QCD in (1+1)-Dimensions,''
  Phys.\ Rev.\ D {\bf 41}, 3814 (1990).


\bibitem{Hellerman:1999nr} 
  S.~Hellerman and J.~Polchinski,
  ``Supersymmetric quantum mechanics from light cone quantization,''
  In Shifman, M.A. (ed.): The many faces of the superworld* 142-155
  [hep-th/9908202].


\bibitem{Polchinski:1999br} 
  J.~Polchinski,
  ``M theory and the light cone,''
  Prog.\ Theor.\ Phys.\ Suppl.\  {\bf 134}, 158 (1999)
  [hep-th/9903165].


\bibitem{Vary:2009gt} 
  J.~P.~Vary, H.~Honkanen, J.~Li, P.~Maris, S.~J.~Brodsky, A.~Harindranath, G.~F.~de Teramond and P.~Sternberg {\it et al.},
  ``Hamiltonian light-front field theory in a basis function approach,''
  Phys.\ Rev.\ C {\bf 81}, 035205 (2010)
  [arXiv:0905.1411 [nucl-th]].


\bibitem{Srivastava:2002mw} 
  P.~P.~Srivastava and S.~J.~Brodsky,
  ``A Unitary and renormalizable theory of the standard model in ghost free light cone gauge,''
  Phys.\ Rev.\ D {\bf 66}, 045019 (2002)
  [hep-ph/0202141].


\bibitem{Brodsky:2009zd} 
  S.~J.~Brodsky and R.~Shrock,
  ``Condensates in Quantum Chromodynamics and the Cosmological Constant,''
  Proc.\ Nat.\ Acad.\ Sci.\  {\bf 108}, 45 (2011)
  [arXiv:0905.1151 [hep-th]].


\bibitem{Brodsky:2012ku} 
  S.~J.~Brodsky, C.~D.~Roberts, R.~Shrock and P.~C.~Tandy,
  ``Confinement contains condensates,''
  Phys.\ Rev.\ C {\bf 85}, 065202 (2012)
  [arXiv:1202.2376 [nucl-th]].


\bibitem{Brodsky:2010xf} 
  S.~J.~Brodsky, C.~D.~Roberts, R.~Shrock and P.~C.~Tandy,
  ``Essence of the vacuum quark condensate,''
  Phys.\ Rev.\ C {\bf 82}, 022201 (2010)
  [arXiv:1005.4610 [nucl-th]].


\bibitem{Brodsky:2006uqa} 
  S.~J.~Brodsky and G.~F.~de Teramond,
  ``Hadronic spectra and light-front wavefunctions in holographic QCD,''
  Phys.\ Rev.\ Lett.\  {\bf 96}, 201601 (2006)
  [hep-ph/0602252].


\bibitem{deTeramond:2008ht} 
  G.~F.~de Teramond and S.~J.~Brodsky,
  ``Light-Front Holography: A First Approximation to QCD,''
  Phys.\ Rev.\ Lett.\  {\bf 102}, 081601 (2009)
  [arXiv:0809.4899 [hep-ph]].


\bibitem{Brodsky:2013ar} 
  S.~J.~Brodsky, G.~F.~De TŽramond and H.~G.~Dosch,
  ``Threefold Complementary Approach to Holographic QCD,''
  Phys.\ Lett.\ B {\bf 729}, 3 (2014)
  [arXiv:1302.4105 [hep-th]].


  
  \bibitem{deAlfaro:1976je}
  V.~de Alfaro, S.~Fubini and G.~Furlan,
  ``Conformal Invariance in Quantum Mechanics,''
Nuovo Cim.\ A {\bf 34}, 569 (1976).



\bibitem{Brodsky:2014yha} 
  S.~J.~Brodsky, G.~F.~de Teramond, H.~G.~Dosch and J.~Erlich,
  ``Light-Front Holographic QCD and Emerging Confinement,''
  arXiv:1407.8131 [hep-ph].


\bibitem{Deur:2014qfa} 
  A.~Deur, S.~J.~Brodsky and G.~F.~de Teramond,
  ``Scheme-Independent Determination of the Perturbative QCD Scale $\bf \Lambda_s$ from Confinement Dynamics in Holographic QCD,''
  arXiv:1409.5488 [hep-ph].


\bibitem{Pauli:1998tf} 
  H.~C.~Pauli,
  ``On confinement in a light cone Hamiltonian for QCD,''
  Eur.\ Phys.\ J.\ C {\bf 7}, 289 (1999)
  [hep-th/9809005].


\bibitem{Glazek:2013jba} 
  S.~D.~Glazek and A.~P.~Trawinski,
  ``Model of the AdS/QFT duality,''
  Phys.\ Rev.\ D {\bf 88}, no. 10, 105025 (2013)
  [arXiv:1307.2059 [hep-ph]].


\bibitem{Appelquist:1977tw} 
  T.~Appelquist, M.~Dine and I.~J.~Muzinich,
  ``The Static Potential in Quantum Chromodynamics,''
  Phys.\ Lett.\ B {\bf 69}, 231 (1977).


\bibitem{Isgur:1984bm} 
  N.~Isgur and J.~E.~Paton,
  ``A Flux Tube Model for Hadrons in QCD,''
  Phys.\ Rev.\ D {\bf 31}, 2910 (1985).


\bibitem{Bjorken:2013boa} 
  J.~D.~Bjorken, S.~J.~Brodsky and A.~Scharff Goldhaber,
  ``Possible multiparticle ridge-like correlations in very high multiplicity proton-proton collisions,''
  Phys.\ Lett.\ B {\bf 726}, 344 (2013)
  [arXiv:1308.1435 [hep-ph]].


\bibitem{Chabysheva:2011ed} 
  S.~S.~Chabysheva and J.~R.~Hiller,
  ``A Light-Front Coupled-Cluster Method for the Nonperturbative Solution of Quantum Field Theories,''
  Phys.\ Lett.\ B {\bf 711}, 417 (2012)
  [arXiv:1103.0037 [hep-ph]].


\bibitem{Trawinski:2014msa} 
  A.~P.~Trawinski, S.~D.~Glazek, S.~J.~Brodsky, G.~F.~de TŽramond and H.~G.~Dosch,
  ``Effective confining potentials for QCD,''
  arXiv:1403.5651 [hep-ph].


\bibitem{Maldacena:1997re} 
  J.~M.~Maldacena,
  ``The Large N limit of superconformal field theories and supergravity,''
  Int.\ J.\ Theor.\ Phys.\  {\bf 38}, 1113 (1999)
  [Adv.\ Theor.\ Math.\ Phys.\  {\bf 2}, 231 (1998)]
  [hep-th/9711200].


\bibitem{Polchinski:2002jw} 
  J.~Polchinski and M.~J.~Strassler,
  ``Deep inelastic scattering and gauge / string duality,''
  JHEP {\bf 0305}, 012 (2003)
  [hep-th/0209211].


\bibitem{Abidin:2008ku} 
  Z.~Abidin and C.~E.~Carlson,
  ``Gravitational form factors of vector mesons in an AdS/QCD model,''
  Phys.\ Rev.\ D {\bf 77}, 095007 (2008)
  [arXiv:0801.3839 [hep-ph]].


\bibitem{Brodsky:2007hb} 
  S.~J.~Brodsky and G.~F.~de Teramond,
  ``Light-Front Dynamics and AdS/QCD Correspondence: The Pion Form Factor in the Space- and Time-Like Regions,''
  Phys.\ Rev.\ D {\bf 77}, 056007 (2008)
  [arXiv:0707.3859 [hep-ph]].


\bibitem{Brodsky:2008pf} 
  S.~J.~Brodsky and G.~F.~de Teramond,
  ``Light-Front Dynamics and AdS/QCD Correspondence: Gravitational Form Factors of Composite Hadrons,''
  Phys.\ Rev.\ D {\bf 78}, 025032 (2008)
  [arXiv:0804.0452 [hep-ph]].


\bibitem{deTeramond:2013it} 
  G.~F.~de Teramond, H.~G.~Dosch and S.~J.~Brodsky,
  ``Kinematical and Dynamical Aspects of Higher-Spin Bound-State Equations in Holographic QCD,''
  Phys.\ Rev.\ D {\bf 87}, no. 7, 075005 (2013)
  [arXiv:1301.1651 [hep-ph]].


\bibitem{deTeramond:2010ge} 
  G.~F.~de Teramond and S.~J.~Brodsky,
  ``Gauge/Gravity Duality and Hadron Physics at the Light-Front,''
  AIP Conf.\ Proc.\  {\bf 1296}, 128 (2010)
  [arXiv:1006.2431 [hep-ph]].


\bibitem{Breitenlohner:1982jf} 
  P.~Breitenlohner and D.~Z.~Freedman,
  ``Stability in Gauged Extended Supergravity,''
  Annals Phys.\  {\bf 144}, 249 (1982).


\bibitem{Karch:2006pv} 
  A.~Karch, E.~Katz, D.~T.~Son and M.~A.~Stephanov,
  ``Linear confinement and AdS/QCD,''
  Phys.\ Rev.\ D {\bf 74}, 015005 (2006)
  [hep-ph/0602229].


\bibitem{Parisi:1972zy} 
  G.~Parisi,
  ``Conformal invariance in perturbation theory,''
  Phys.\ Lett.\ B {\bf 39}, 643 (1972).


\bibitem{Leutwyler:1977vy} 
  H.~Leutwyler and J.~Stern,
  ``Relativistic Dynamics on a Null Plane,''
  Annals Phys.\  {\bf 112}, 94 (1978).


\bibitem{deTeramond:2014yga} 
  G.~F.~de TŽramond, S.~J.~Brodsky and H.~G.~Dosch,
  ``Hadron Spectroscopy and Dynamics from Light-Front Holography and Conformal Symmetry,''
  EPJ Web Conf.\  {\bf 73}, 01014 (2014)
  [arXiv:1401.5531 [hep-ph]].


\bibitem{deTeramond:2014rsa} 
  G.~F.~de Teramond, S.~J.~Brodsky and H.~G.~Dosch,
  ``Light-Front Holography in QCD and Hadronic Physics,''
  arXiv:1405.2451 [hep-ph].


\bibitem{deTeramond:2012rt} 
  G.~F.~de Teramond and S.~J.~Brodsky,
  ``Hadronic Form Factor Models and Spectroscopy Within the Gauge/Gravity Correspondence,''
  arXiv:1203.4025 [hep-ph].


\bibitem{Brodsky:2011xx} 
  S.~J.~Brodsky, F.~G.~Cao and G.~F.~de Teramond,
  ``Meson Transition Form Factors in Light-Front Holographic QCD,''
  Phys.\ Rev.\ D {\bf 84}, 075012 (2011)
  [arXiv:1105.3999 [hep-ph]].


\bibitem{deTeramond:2011qp} 
  G.~F.~de Teramond and S.~J.~Brodsky,
  ``Excited Baryons in Holographic QCD,''
  AIP Conf.\ Proc.\  {\bf 1432}, 168 (2012)
  [arXiv:1108.0965 [hep-ph]].


\bibitem{Forshaw:2012im} 
  J.~R.~Forshaw and R.~Sandapen,
  ``An AdS/QCD holographic wavefunction for the rho meson and diffractive rho meson electroproduction,''
  Phys.\ Rev.\ Lett.\  {\bf 109}, 081601 (2012)
  [arXiv:1203.6088 [hep-ph]].


\bibitem{Vega:2010ns} 
  A.~Vega, I.~Schmidt, T.~Gutsche and V.~E.~Lyubovitskij,
  ``Generalized parton distributions in AdS/QCD,''
  Phys.\ Rev.\ D {\bf 83}, 036001 (2011)
  [arXiv:1010.2815 [hep-ph]].


\bibitem{Chakrabarti:2013dda} 
  D.~Chakrabarti and C.~Mondal,
  ``Nucleon and flavor form factors in a light front quark model in AdS/QCD,''
  Eur.\ Phys.\ J.\ C {\bf 73}, 2671 (2013)
  [arXiv:1307.7995 [hep-ph]].


\bibitem{Weinberg:1967kj} 
  S.~Weinberg,
  ``Precise relations between the spectra of vector and axial vector mesons,''
  Phys.\ Rev.\ Lett.\  {\bf 18}, 507 (1967).


\bibitem{Shifman:1978bx} 
  M.~A.~Shifman, A.~I.~Vainshtein and V.~I.~Zakharov,
  ``QCD and Resonance Physics. Sum Rules,''
  Nucl.\ Phys.\ B {\bf 147}, 385 (1979).


\bibitem{Grunberg:1980ja} 
  G.~Grunberg,
  ``Renormalization Group Improved Perturbative QCD,''
  Phys.\ Lett.\ B {\bf 95}, 70 (1980)
  [Erratum-ibid.\ B {\bf 110}, 501 (1982)].


\bibitem{Cvetic:2013gta} 
  G.~Cvetic,
  ``Techniques of evaluation of QCD low-energy physical quantities with running coupling with infrared fixed point,''
  Phys.\ Rev.\ D {\bf 89}, 036003 (2014)
  [arXiv:1309.1696 [hep-ph]].


\bibitem{Cornwall:1981zr} 
  J.~M.~Cornwall,
  ``Dynamical Mass Generation in Continuum QCD,''
  Phys.\ Rev.\ D {\bf 26}, 1453 (1982).


\bibitem{Shirkov:1997wi} 
  D.~V.~Shirkov and I.~L.~Solovtsov,
  ``Analytic model for the QCD running coupling with universal alpha-s (0) value,''
  Phys.\ Rev.\ Lett.\  {\bf 79}, 1209 (1997)
  [hep-ph/9704333].


\bibitem{Fischer:2003rp} 
  C.~S.~Fischer and R.~Alkofer,
  ``Nonperturbative propagators, running coupling and dynamical quark mass of Landau gauge QCD,''
  Phys.\ Rev.\ D {\bf 67}, 094020 (2003)
  [hep-ph/0301094].


\bibitem{Aguilar:2009nf} 
  A.~C.~Aguilar, D.~Binosi, J.~Papavassiliou and J.~Rodriguez-Quintero,
  ``Non-perturbative comparison of QCD effective charges,''
  Phys.\ Rev.\ D {\bf 80}, 085018 (2009)
  [arXiv:0906.2633 [hep-ph]].


\bibitem{Courtoy:2013apa} 
  A.~Courtoy and S.~Liuti,
  ``Analysis of $\alpha_s$ from the realization of quark-hadron duality,''
  Int.\ J.\ Mod.\ Phys.\ Conf.\ Ser.\  {\bf 25}, 1460046 (2014)
  [arXiv:1307.4211].


\bibitem{Bjorken:1966jh} 
  J.~D.~Bjorken,
  ``Applications of the Chiral U(6) x (6) Algebra of Current Densities,''
  Phys.\ Rev.\  {\bf 148}, 1467 (1966).


\bibitem{Deur:2005cf} 
  A.~Deur, V.~Burkert, J.~P.~Chen and W.~Korsch,
  ``Experimental determination of the effective strong coupling constant,''
  Phys.\ Lett.\ B {\bf 650}, 244 (2007)
  [hep-ph/0509113].
  
\bibitem{Deur:2008rf} 
  A.~Deur, V.~Burkert, J.~P.~Chen and W.~Korsch,
  ``Determination of the effective strong coupling constant alpha(s,g(1))(Q**2) from CLAS spin structure function data,''
  Phys.\ Lett.\ B {\bf 665}, 349 (2008)
  [arXiv:0803.4119 [hep-ph]].


\bibitem{Drell:1966jv} 
  S.~D.~Drell and A.~C.~Hearn,
  ``Exact Sum Rule for Nucleon Magnetic Moments,''
  Phys.\ Rev.\ Lett.\  {\bf 16}, 908 (1966).


\bibitem{Gerasimov:1965et} 
  S.~B.~Gerasimov,
  ``A Sum rule for magnetic moments and the damping of the nucleon magnetic moment in nuclei,''
  Sov.\ J.\ Nucl.\ Phys.\  {\bf 2}, 430 (1966)
  [Yad.\ Fiz.\  {\bf 2}, 598 (1965)].


\bibitem{Brodsky:2010ur} 
  S.~J.~Brodsky, G.~F.~de Teramond and A.~Deur,
  ``Nonperturbative QCD Coupling and its $\beta$-function from Light-Front Holography,''
  Phys.\ Rev.\ D {\bf 81}, 096010 (2010)
  [arXiv:1002.3948 [hep-ph]].


\bibitem{Brodsky:1994eh} 
  S.~J.~Brodsky and H.~J.~Lu,
  ``Commensurate scale relations in quantum chromodynamics,''
  Phys.\ Rev.\ D {\bf 51}, 3652 (1995)
  [hep-ph/9405218].


\bibitem{Brodsky:1995tb} 
  S.~J.~Brodsky, G.~T.~Gabadadze, A.~L.~Kataev and H.~J.~Lu,
  ``The Generalized Crewther relation in QCD and its experimental consequences,''
  Phys.\ Lett.\ B {\bf 372}, 133 (1996)
  [hep-ph/9512367].


\bibitem{Wu:2013ei} 
  X.~G.~Wu, S.~J.~Brodsky and M.~Mojaza,
  ``The Renormalization Scale-Setting Problem in QCD,''
  Prog.\ Part.\ Nucl.\ Phys.\  {\bf 72}, 44 (2013)
  [arXiv:1302.0599 [hep-ph]].


\bibitem{Deur:2004ti} 
  A.~Deur, {\it et al.},
  ``Experimental determination of the evolution of the Bjorken integral at low Q**2,''
  Phys.\ Rev.\ Lett.\  {\bf 93}, 212001 (2004)
  [hep-ex/0407007].
  
\bibitem{Deur:2014vea} 
  A.~Deur, Y.~Prok, V.~Burkert, D.~Crabb, F.-X.~Girod, K.~A.~Griffioen, N.~Guler and S.~E.~Kuhn {\it et al.},
  ``High precision determination of the $Q^2$-evolution of the Bjorken Sum,''
  Phys.\ Rev.\ D {\bf 90}, 012009 (2014)
  [arXiv:1405.7854 [nucl-ex]].


\bibitem{Bloom:1970xb} 
  E.~D.~Bloom and F.~J.~Gilman,
  ``Scaling, Duality, and the Behavior of Resonances in Inelastic electron-Proton Scattering,''
  Phys.\ Rev.\ Lett.\  {\bf 25}, 1140 (1970).


\bibitem{Kataev:1994gd} 
  A.~L.~Kataev,
  ``The Ellis-Jaffe sum rule: The Estimates of the next to next-to-leading order QCD corrections,''
  Phys.\ Rev.\ D {\bf 50}, 5469 (1994)
  [hep-ph/9408248].


\bibitem{Lepage:1991zt} 
  G.~P.~Lepage and P.~B.~Mackenzie,
  ``Renormalized lattice perturbation theory,''
  Nucl.\ Phys.\ Proc.\ Suppl.\  {\bf 20}, 173 (1991).


\bibitem{Davies:2003ik} 
  C.~T.~H.~Davies {\it et al.}  [HPQCD and UKQCD and MILC and Fermilab Lattice Collaborations],
  ``High precision lattice QCD confronts experiment,''
  Phys.\ Rev.\ Lett.\  {\bf 92}, 022001 (2004)
  [hep-lat/0304004].


 
\end{thebibliography}
\end{document}